# A portable X-pinch design for x-ray diagnostics of warm dense matter

J. Strucka, J. W. D. Halliday, T. Gheorghiu, et al.

# A portable X-pinch design for x-ray diagnostics of warm dense matter

View Online Export Citation CrossMark

J. Strucka,[1,a)] J. W. D. Halliday,[1] T. Gheorghiu,[1] H. Horton,[2] B. Krawczyk,[1] P. Moloney,[1] S. Parker,[1] G. Rowland,[1] N. Schwartz,[3] S. Stanislaus,[1] S. Theocharous,[1] C. Wilson,[1] Z. Zhao,[1] T. A. Shelkovenko,[4] S. A. Pikuz,[4] and S. N. Bland[1]

## AFFILIATIONS

[1]Plasma Physics Group, Imperial College London, London SW7 2BW, United Kingdom
[2]University of Cambridge, Cambridge, United Kingdom
[3]Department of Nuclear, Plasma, and Radiological Engineering, University of Illinois at Urbana-Champaign, Urbana, Illinois 61801, USA
[4]Lebedev Physical Institute, Russian Academy of Sciences, Moscow 119991, Russia

[a)]**Author to whom correspondence should be addressed:** jergus.strucka15@imperial.ac.uk

## ABSTRACT

We describe the design and x-ray emission properties (temporal, spatial, and spectral) of Dry Pinch I, a portable X-pinch driver developed at Imperial College London. Dry Pinch I is a direct capacitor discharge device, $300 \times 300 \times 700$ mm$^3$ in size and ~50 kg in mass, that can be used as an external driver for x-ray diagnostics in high-energy-density physics experiments. Among key findings, the device is shown to reliably produce $1.1 \pm 0.3$ ns long x-ray bursts that couple ~50 mJ of energy into photon energies from 1 to 10 keV. The average shot-to-shot jitter of these bursts is found to be $10 \pm 4.6$ ns using a combination of x-ray and current diagnostics. The spatial extent of the x-ray hot spot from which the radiation emanates agrees with previously published results for X-pinches—suggesting a spot size of $10 \pm 6$ $\mu$m in the soft energy region (1–10 keV) and $190 \pm 100$ $\mu$m in the hard energy region ( >10 keV). These characteristics mean that Dry Pinch I is ideally suited for use as a probe in experiments driven in the laboratory or at external facilities when more conventional sources of probing radiation are not available. At the same time, this is also the first detailed investigation of an X-pinch operating reliably at current rise rates of less than 1 kA/ns.

## I. INTRODUCTION

Research into warm dense matter represents a frontier in our understanding of material physics. Such matter is characterized by temperatures comparable to the Fermi energy and a coupling parameter close to unity. Here condensed matter theory crosses into plasma physics, and models are often incomplete. Condensed matter approaches such as Kohn–Sham density functional theory are not computationally feasible,[1] and classical plasma physics approximations are no longer valid. The lack of predictive capabilities highlights the importance of experimental results to aid the development of theoretical and computational models. Dynamic compression facilities capable of producing warm dense matter using pulsed power approaches are widely available in the laboratory setting, but often suffer from lack of access to suitable diagnostics. X-ray diagnostics, in particular, are an important tool in understanding states of matter inaccessible to optical radiation, such as optically thick materials produced in warm dense matter research[2] or inertial confinement fusion schemes.[3]

Available x-ray diagnostic methods include point-projection radiography for imaging of highly absorptive interfaces,[4] phase-contrast imaging for interfaces with negligible changes in absorption,[5,6] x-ray diffraction to probe the internal geometrical structure of novel material states,[7] and x-ray absorption fine structure spectroscopy to provide information about the temperature, density,[8] or electronic structure of materials.[9] These diagnostics can be driven by large high-energy laser–matter interaction experiments at national facilities[8] such as the Laboratory for Laser Energetics (LLE) and the Rutherford Appleton Laboratory (RAL). In some cases, experiments can also be temporarily relocated to take advantage of third-generation synchrotrons[9,10] or x-ray free-electron lasers (XFELs);

however, in many cases, this option is not available. Further, even for experiments where relocation to external facilities is possible, there is a considerable benefit in having on-site x-ray diagnostics to obtain preliminary data and optimize the experiment prior to facility beam time.

X-pinches are relatively novel devices derived from pulsed power-driven Z-pinch research.[11,12] In its simplest form, a Z-pinch consists of a single metallic wire subject to a fast and high current pulse (hundreds of kiloamperes in hundreds of nanoseconds). The wire is resistively heated and forms a plasma. Simultaneously, the current compresses ("pinches") the wire along its length owing to the $\mathbf{J} \times \mathbf{B}$ force exerted on the moving charge carriers. Equilibrium compression, when the plasma pressure equals the magnetic pressure, is an unstable equilibrium, and any small-scale perturbations along the wire rapidly grow in what is known as the $m = 0$ (or sausage) instability. As a result of this instability, a number of local points are compressed to high pressures and temperatures—these regions are called hot spots and they radiate extreme amounts of radiation. The location of the hot spots is randomly seeded and cannot be predicted before the experiment. In 1979, research on the Don facility at the Lebedev Physical Institute, USSR pioneered the use of noncylindrical wire loads—wires crossed to form an "X"—to localize the hot-spot formation by concentrating the magnetic field, thus enhancing the instability growth rate at the crossing point.[13]

In general, an X-pinch is formed by wires that are a few micrometers thick and are crossed to form an "X." Typically, a current pulse of tens to hundreds of kiloamperes is passed through the wire load with a rise faster than 1 kA/ns to reliably produce a hot spot. This empirical observation, sometimes referred to as the Shelkovenko condition, has historically separated X-pinches in terms of reliability and emission. The x-ray emission from an X-pinch is produced by one of two physical processes co-occurring within the plasma. During the collapse of the wire plasma, soft thermal emission ($E \sim$ 1–10 keV) from the highly compressed and heated region (soft hot spot) radiates a short (tens of picoseconds) and intense burst of x-ray radiation with good imaging properties (source size $\sim 5\ \mu$m).[14] Immediately following the compression, an electron cascade accelerated across the anode–cathode (A–K) gap hits the dense plasma mini-diode, yielding harder ($E >$ 10 keV) bremsstrahlung radiation over multiple nanoseconds (hard hot spot) with a typical source size $>50\ \mu$m. Spectral features of the radiation produced by an X-pinch include strong emission lines that can be used for x-ray diffraction measurements[15] and smooth broadband continua—due to the intense pressures and densities reached in the hot spot—that can be exploited for absorption spectroscopy.[16]

Until recently, high-current pulsed power systems were typically designed around Marx banks (oil-filled reservoirs with capacitors connected by $SF_6$-filled switches) and pulse-sharpening transmission lines filled with deionized water. Recent advances in the design of high-voltage components, such as low-impedance capacitors and reliable dry air switches for linear transformer driver (LTD) technology, have enabled the development of more easily portable X-pinch drivers, including an oil, water, and $SF_6$-free X-pinch driver at the Idaho Accelerator Centre,[17] as well as the MINI and KING facilities at the Lebedev Institute.[18] However, none of these drivers have been designed or utilized for probing experiments performed external to the X-pinch driver—such as on a second pulsed power system, laser-produced plasma, or gas gun experiment. This places strict criteria on the jitter of the X-pinch driver, its photon flux, and its reliability.

In this paper, we describe the reliability and radiation performance of Dry Pinch I—a portable direct capacitor discharge X-pinch device developed at Imperial College London in 2016 for use in probing external experiments. The remainder of the paper is structured as follows: in Sec. II, we describe the technical specifications and pulsed power performance of Dry Pinch I, in Sec. III, we describe the diagnostics used for the work presented in this paper, in Sec. IV, we summarize the experimental findings and x-ray emission performance of the driver (in both the soft and hard x-ray regimes), and finally, in Sec. V, we summarize the results and discuss future work.

## II. TECHNICAL SPECIFICATIONS

Dry Pinch I—depicted in Fig. 1—is a novel X-pinch driver developed at Imperial College London. It is a small portable X-pinch ~50 kg in weight and 300 × 300 × 700 $mm^3$ in size. Unlike other X-pinches, its design is focused on ease of use and quick adaptability to larger experimental platforms, rather than absolute optimization of inductance. It includes a long (usually ~100 mm) insertion stalk to use inside external experimental vacuum chambers as a source for x-ray diagnostics.

Dry Pinch I is a direct capacitor discharge driver that can be thought of as a serial connection of a closing switch, a capacitor, an inductance, and a load. An equivalent electrical circuit representing the driver with a short-circuited load is shown in Fig. 2. Following the numbering convention from Fig. 1, the driver is operated as follows: an external high-voltage supply with positive polarity charges the capacitors in parallel via cables connected through an entry port in the aluminum casing (1). Once the capacitors (2) reach the positive charging voltage $V$, an external trigger pulse is coupled into each of the switches (3) and starts a controlled breakdown, effectively closing the switches and connecting the high-voltage side of the capacitors to ground. The voltage on the other side of the capacitor flips to $-V$ and discharges through the load. The capacitors inside the casing are connected in parallel by a transfer plate (4) that is separated from the grounded aluminum casing by a 12 mm thick high-density polyethylene insulating plate (5) cut at a 45° angle to increase the breakdown voltage. Upon closing of the switches, the capacitors discharge in parallel through a converging conical section (6) into the magnetically insulating transmission line (MITL) (7) with a non-integrating Rogowski coil placed within the casing to measure the current flow through the transmission line (8). The current is then guided through a gently converging cathode (9) and a load locking ring (10), through the X-pinch load (11), and straight into the anode (12), which is connected to ground via the external casing.

The work of Shapovalov *et al.*[17] at the Idaho Accelerator Center and Dry Pinch I are both based on a brick design developed for the MACH generator at Imperial College London by Dr. Spielman at the Ktech Corporation. The capacitors within the casing are packaged as two side-by-side LTD bricks. Encased in monolithic resin, the inside of each brick contains two double-ended 150 nF low-inductance General Atomics capacitors (Part 35465) rated up to 100 kV and a single multichannel and multigap ball switch [depicted in Fig. 1(d)] operated in dry air at atmospheric pressure. Each switch has five channels separated by 20 mm and seven 5.6 mm gaps. Assuming a static breakdown voltage in dry air, each gap in the switch is rated up to 16.8 kV, with the entire switch theoretically capable of holding ~120 kV. Prior to triggering, the operating voltage is uniformly distributed between the rows of balls by a chain of highly resistive resistors until a trigger pulse perturbs the voltage distribution, ionizes the air between the first and second rows, and leads to a controlled breakdown of the channels.

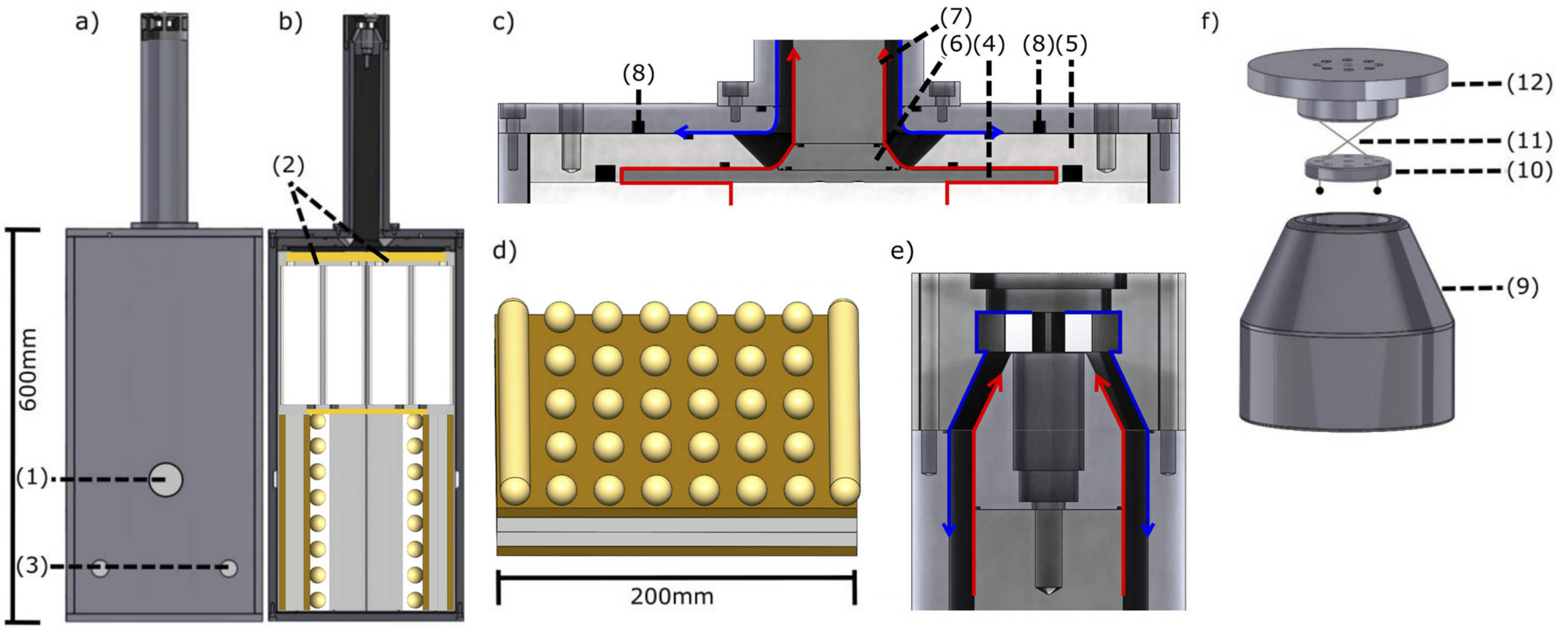

**FIG. 1.** (a) CAD model of the entire external housing with a 350 mm long transmission line together with a circular entry port (1) for high-voltage charging cables and small circular entry ports (3) for triggering cables coupled into the switches. (b) Cross-section of Dry Pinch I containing the LTD bricks (2). The capacitors and switches are not to scale. (c) Close-up of the transfer plate zone, including the transfer plate connecting the two capacitors (4), a high-density polyethylene insulating plate (5), a gently converging current plate (6), a magnetically insulating transmission line (7), and a recess for a nonintegrating Rogowski coil (8). (d) CAD model of the five-channel dry air ball gap switch used to trigger Dry Pinch I. (e) Cross-section of the final part of the transmission line where the X-pinch load is mounted. The gap located in the cathode at the center of the diagram hosts the load-locking ring [shown in (f)]. It is possible to use different designs for the load section with different A–K gaps or shapes/numbers of output windows. (f) CAD model of the mounting mechanism, including the cathode (9), the load-locking ring (10), the X-pinch load (11), and the top anode plug (12). The direction of electron flow within the device is indicated by red arrows before the load and by blue arrows after the load.

At present, charging of the capacitors is handled by an external Glassman series WS power supply through 150 MΩ resistors. The resistors are in place to limit the charging current and prevent any issues between the two bricks that could be caused by fast, non-uniform charging. The high-voltage power supply is connected to the brick charging resistors through a safety system consisting of a diode stack submerged in a barrel of transformer oil. The safety system is in place to prevent any damage to the power supply due to voltage reversals and ringing within the system. The safety system also includes a pneumatically operated voltage dump that connects the capacitors directly to ground inside the barrel. Charging voltage is usually limited to 85 kV to slow down deterioration of the capacitor performance over many shots due to current reversals.

Once the system is charged, a trigger voltage is provided by an external TG75 vector inversion generator that capacitatively couples to the second row of balls in the switch via a stripped RG214 coaxial cable. It delivers a 75 kV trigger pulse into an open-ended transmission line that doubles the pulse to 150 kV. All of the external components, including the charging, safety, and trigger systems, are mounted on a rack that can be located at a distance from the X-pinch.

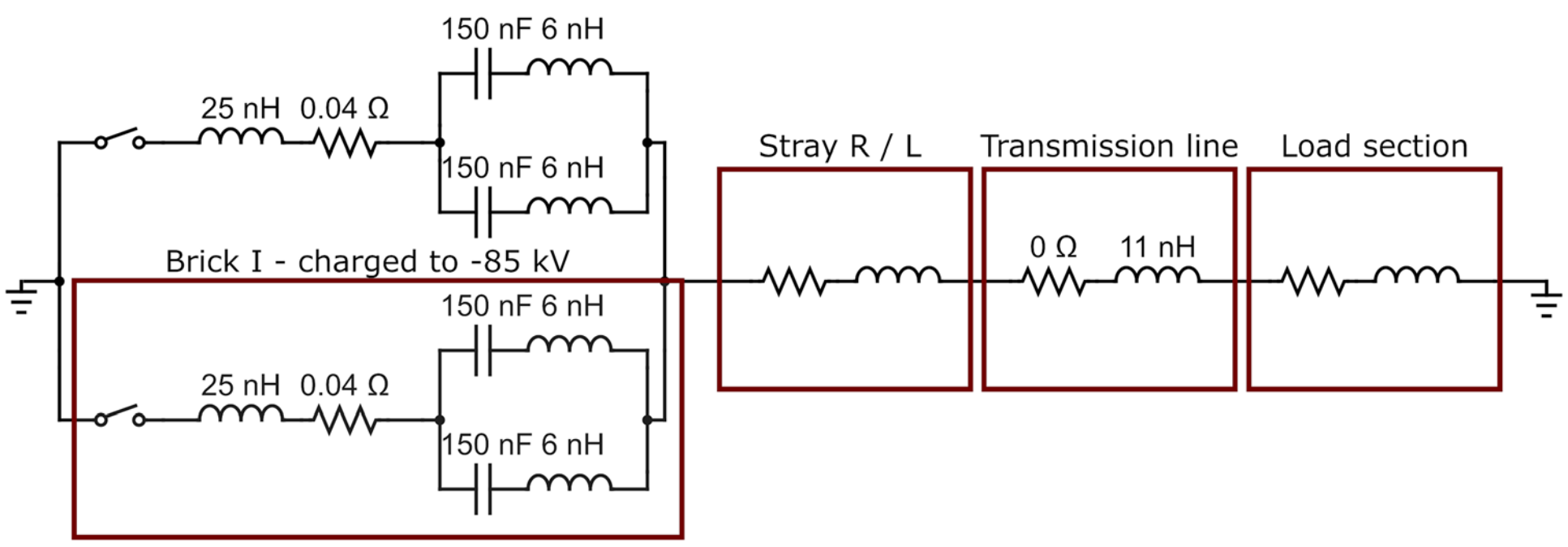

**FIG. 2.** Equivalent circuit diagram of the X-pinch and its respective components. The inductance of the MITL was modeled as arising from two coaxial cylinders discharging in parallel.

In terms of mechanical design, all elements of the system, with the exception of the plastic insulation plate (5) and brass anode plug (12), are fabricated from aluminum. The inner and outer diameters of the coaxial transmission line are 44 and 50.8 mm, respectively. To ensure smooth operation of the device and prevent unexpected breakdowns, it is important to regularly bead-blast all surfaces that are facing high-potential gradients, especially within the magnetically insulating transmission line. The reliability of the switches is improved by purging them with a continuous supply of dry air for at least 20 min before the first shot of the day.

Figure 3 shows a current waveform produced by Dry Pinch I with the capacitors charged to 85 kV and discharged through a short-circuited load consisting of an M6 bolt in the A–K gap. The system reaches a maximum current of 140 kA in 350 ns, with a 10%–90% rise rate of 0.5 kA/ns. These measurements are likely an underestimate of the actual peak current and current rise time, since the calibration of the nonintegrating Rogowski coils has shown a discrepancy of ~20% between standardized Rogowski coils and the coil enclosed within the metal recess of the X-pinch case.

To date, the reliability of the pulsed power system has been tested in over 100 shots using a multitude of common loads ranging from wire loads to gas puffs. Figure 7(b) shows the current waveforms of 17 back-to-back shots with a four-wire load and a charging voltage of 85 kV. During the series of experiments, the pulsed power system self-triggered twice near maximum voltage. In successful shots, on average, the current peak is achieved at 127 kA with a 0%–100% rise time of 350 ns and a 10%–90% rise time of 230 ns. The average current rise rate of 0.45 kA/ns is below the empirical Shelkovenko condition usually required to achieve reliable hot spot formation in X-pinches; however, the x-ray emission results presented in Sec. IV prove that Dry Pinch I is reliable in spite of this. In all experiments presented in this paper, the length of the insertion stalk was set at 100 mm.

## III. DIAGNOSTICS

Owing to its compact size, a design optimized for compatibility with most experimental chambers, and isotropic 4$\pi$ emission, Dry Pinch I can be simultaneously equipped with a wide array of optical, x-ray, and electronic diagnostics to evaluate its performance. The x-ray diagnostics employed in our experiments can be further split into three separate categories: spatially, temporally, and spectrally resolved diagnostics.

Optical diagnostics were utilized to characterize the overall dynamics of the X-pinch from plasma formation at the wires, throughout the pinching process, and at eventual breakup. A system of perfect telescopes was used to transmit the light emitted by the plasma to an optical fast framing camera (Invisible Vision UHSi) capable of recording 12 frames with 20 ns interframe time and 5 ns exposure. Because the optical emission from the X-pinch is too intense and would saturate the camera, it was filtered by a bandpass filter centered at $\lambda_c$ = 532 nm with a transmission full width at half maximum (FWHM) of 1 nm.

Spatially resolved x-ray images were recorded on Fuji BAS-IP MS 2040 E image plates and Kodak Direct Exposure Film (DEF). The Fuji BAS-IP image plates are sensitive to a wide range of x-ray energies between 1 and 90 keV, with peak sensitivity at 18 keV. The Kodak DEF is selectively sensitive to soft x-rays with peak sensitivity at 4 keV. Figure 4 shows the sensitivity curves for these media as measured and modeled by Meadowcroft *et al.*[19] and Henke *et al.*[20] These recording media were used in conjunction to decouple the soft thermal emission from the hard radiation component by stacking the DEF on top of a Fuji image plate.

X-ray pinhole imaging was usually fielded in two configurations that depended on the respective sizes of the source and the pinhole. If the characteristic source size $r_s$ was much larger than the pinhole radius $r_p$, the pinhole effectively formed a camera obscura with a magnification $M = d_s/d_i$, where $d_s$ is the distance from the source to the pinhole and $d_i$ is the distance from the pinhole to the image plane. If $r_s$ is much smaller than $r_p$, then the source projects the shape of the pinhole onto the image plane with magnification $M = (d_s + d_i)/d_s$ and typical resolution $(M - 1)r_s$. Because the soft x-ray hot spot produced by X-pinches is characteristically a few micrometers in size, approximately equal to the size of the smallest transmission-limited pinholes, the spatial diagnostics presented in this paper worked in the large-pinhole regime.

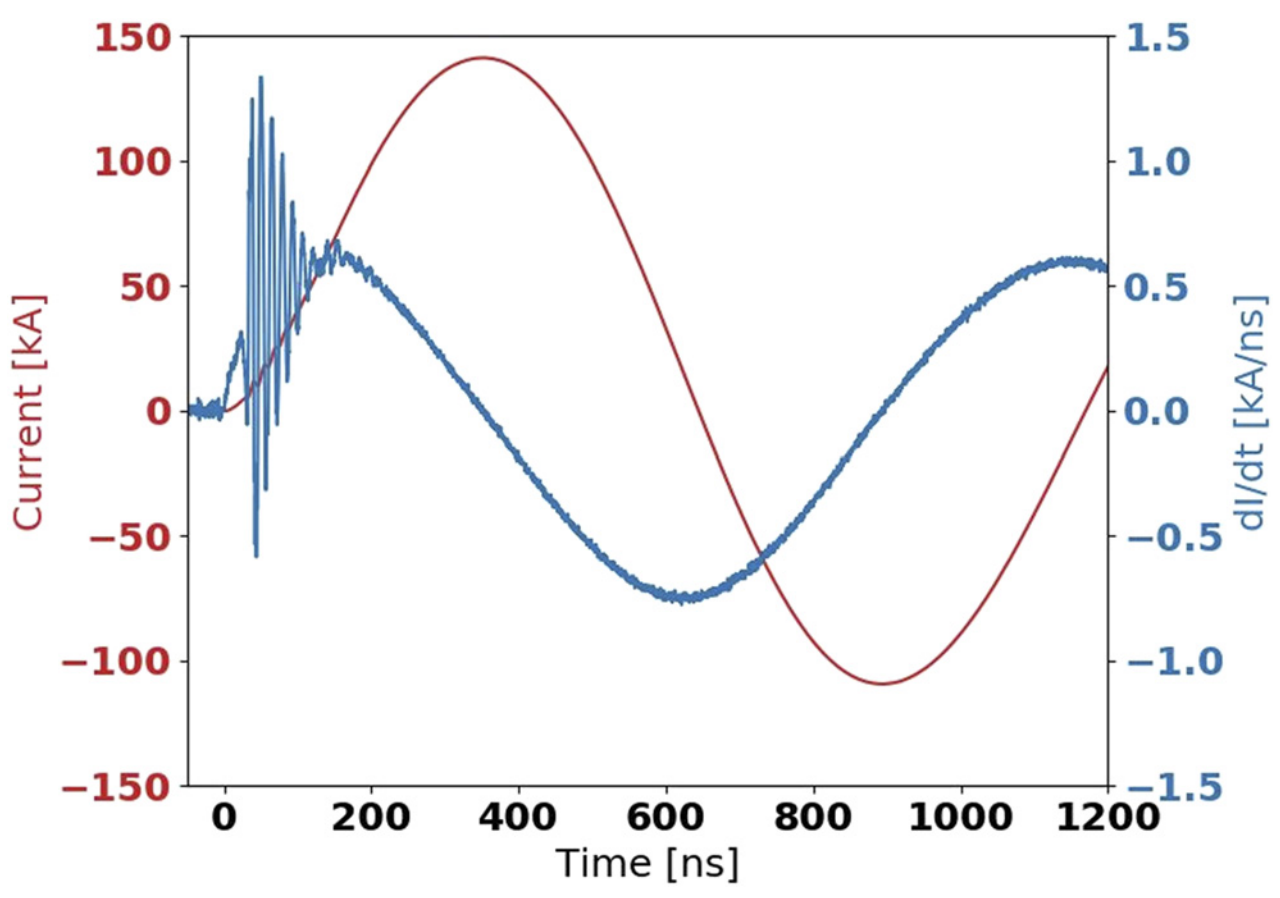

**FIG. 3.** Short-circuit current measurement with the load replaced by an M6 bolt. Nonintegrated Rogowski data were measured by a coil located within the recess of the Dry Pinch I top plate (depicted in blue). The oscillatory signal observed at the beginning of the Rogowski data is the noise caused by the switch firing. The red line corresponds to the current calculated as a cumulative integral of the nonintegrated Rogowski data.

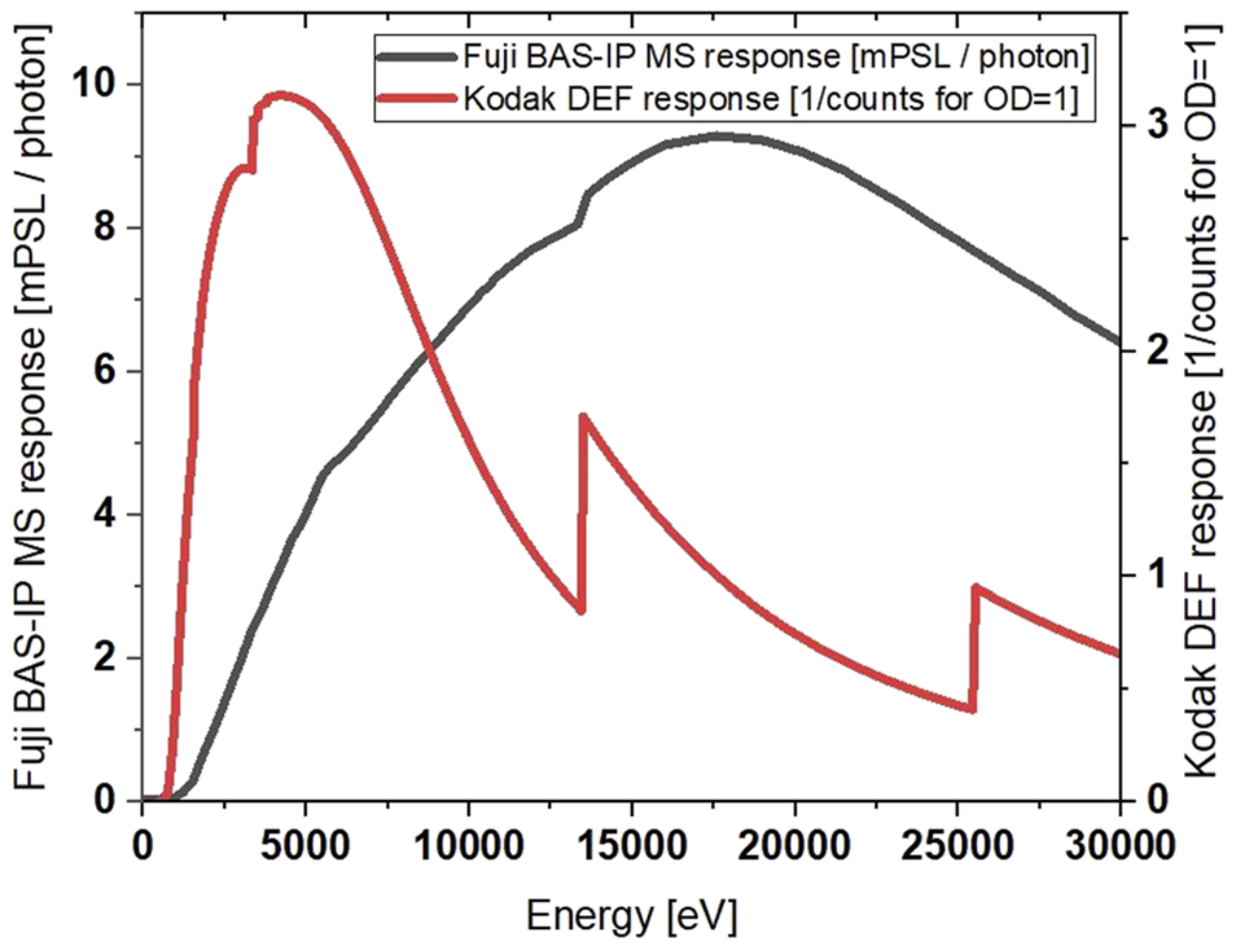

**FIG. 4.** The x-ray sensitivity of the Fuji-IP MS image plate (plotted in black) and the sensitivity of the Kodak DEF (plotted in red). Data are taken from experiments and modeling by Meadowcroft *et al.*[19] and Henke *et al.*[20]

To explore point-projection radiography, we measured the transmission through an array of (nominally) 200 $\mu$m pinholes laser-cut in a 100 ± 10 $\mu$m thick circular stainless steel substrate with a diameter of 40 mm. To investigate source sizes for x-rays with different spectral properties, the array was overlaid with multiple metallic foils. These simulated objects with particular spectral transmission profiles (absorption of soft x-rays). Transmission profiles corresponding to the combinations of the metallic foils attached to the pinhole array are plotted in Fig. 5(a). Other radiography targets included a 100 $\mu$m thick stainless steel 1951 United States Air Force (USAF) resolution chart with elements down to 35 $\mu$m in size.[21]

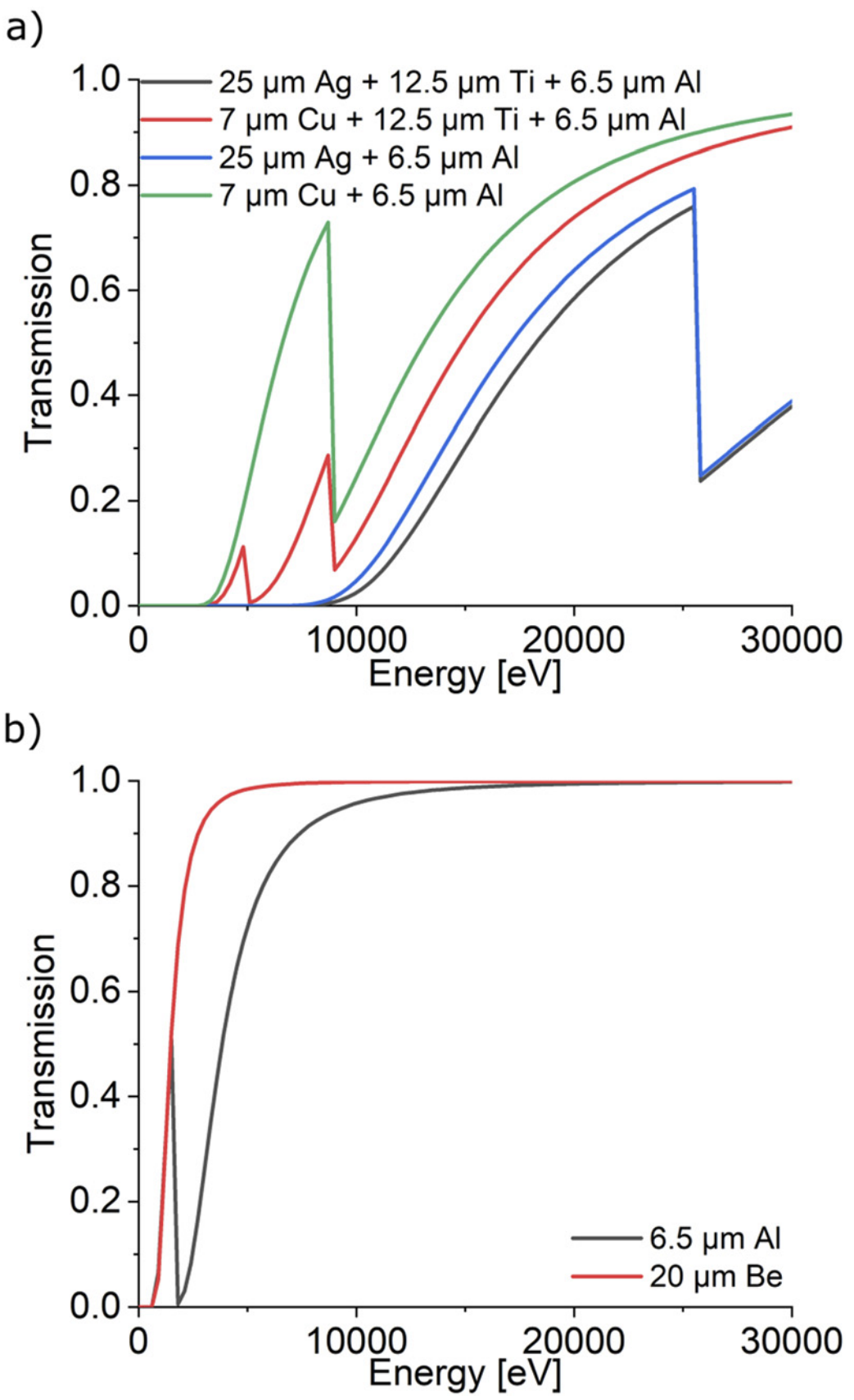

**FIG. 5.** Transmission profile of metal filters applied to (a) a stainless steel pinhole array target and (b) diamond photoconductive detectors and silicon x-ray diodes.

Diagnostics with temporal resolution included two diamond photoconducting detectors (PCDs) with a characteristic electronic recombination time of 120 ps. These were connected (via 50 Ω terminating resistors) to oscilloscope channels that had a 200 ps sampling period. Combining these two timescales in quadrature yields a temporal resolution of 230 ps. The size of the sensitive element within the PCD was $1 \times 3 \times 0.5$ mm$^3$ (W × L × D). Calibration of equivalent detectors by Sandia National Laboratories using a synchrotron yielded a typical response of $6 \times 10^{-4}$ A/W at a bias voltage of 100 V.[22] The response function of the detectors was also found to be virtually flat up to 3 keV, with a drop at higher energies. In our experiments, a set of PCDs with the same design was operated at 350 V bias voltage. The sensitivity was then assumed to be $2.1 \times 10^{-3}$ A/W throughout the analysis. Silicon x-ray photodiodes (AXUV SiD) were also fielded for their higher sensitivity to harder x-rays (up to 50 keV). To alter their spectral response, these detectors were filtered using either 20 $\mu$m thick beryllium or 6.5 $\mu$m thick aluminum. Transmission curves for these filters are plotted in Fig. 5(b). These filters were chosen for their similar transmission profiles below the aluminum K-edge at 1.5 keV and above 10 keV, with the aluminum filter strongly suppressing radiation between these two thresholds. Comparatively, the sensitivity of the beryllium filtered diodes is therefore strongly enhanced in the energy region where thermal emission dominates.

Spectrally resolved diagnostics consisted of a cylindrical (convex) lithium fluoride spectrometer with a lattice spacing of $2d = 0.402$ nm and a crystal radius of $r_c = 150$ mm. Design of the spectrometer followed Henke *et al.*,[23] with the recording image plate placed on a circle of 83 mm radius and centered at the virtual focal point of the crystal. The crystal was protected from X-pinch debris and low-energy radiation either by 12.5 $\mu$m thick titanium or 6.5 $\mu$m thick aluminum foil placed between the X-pinch and the crystal. The energy bandwidth of the spectrometer spanned the region from 7 to 36 keV, corresponding to a Bragg angle of $\theta = 15^\circ$ for the central ray, or from 6 to 18.5 keV for a Bragg angle of $\theta = 20^\circ$.

Lastly, an aluminum step-wedge with thickness from 100 to 1500 $\mu$m in increments of 100 $\mu$m was used to calculate the "bremsstrahlung temperature" of the X-pinch emission. The signal transmitted through the steps was plotted as a function of thickness, and was fitted by a signal corresponding to theoretical transmission for a source with spectral intensity given by

$$I(A, E, T) = A \exp(-E/T), \tag{1}$$

where $E$ is the x-ray energy in units of eV, $T$ is the "bremsstrahlung temperature" in eV, and $A$ is a normalization parameter. This approach can be used to establish the importance of continuum emission compared with line emission by observing the relationship between the signal value and the step thickness. It is also a simple benchmark that can be easily fielded on most X-pinches to facilitate a qualitative comparison of the spectra produced by different generators.

## IV. DRY PINCH I PERFORMANCE: RESULTS AND DISCUSSION

Dry Pinch I was used as a driver for four-wire X-pinches. The wire material and diameter were chosen to optimize the coupling of electrical energy into x-ray radiation. Theoretically, the linear density of the wires should be chosen such that the X-pinch forms approximately at the peak current; however, experiments show that optimal pinching occurs if the X-pinch forms between the moment of peak current and that of peak current derivative.[14] In our experience, optimizing for the latter of these two conditions produced the best x-ray yield and operational reliability.

Multiple materials were fielded, with an A–K gap of 10 mm and a crossing angle of $\theta = 90^{\circ}$. The wire materials that produced a signal on the x-ray diodes included silver, gold, molybdenum, tungsten, and nickel–chromium. Silver X-pinches yielded the most reliable imaging properties and were tested further for use in radiography and other experimental techniques.

Optical images captured by the fast framing camera shown in Fig. 6 confirm that the silver load undergoes initial thermal expansion and subsequent compression due to magnetic pressure, as expected for an X-pinch load.

## A. Reliability

A wire load made of four silver wires—each 30 $\mu$m in diameter ($\rho_{\mathrm{lin}}$ = 59.3 mg/m)—was used in 17 back-to-back shots to obtain a representative sample for the reliability and imaging performance of the X-pinch. In every shot, the capacitors were charged up to 85 kV. Current waveforms obtained from the nonintegrated Rogowski coil located around the magnetically insulated transmission line and filtered by a 100-point (20 ns) moving-average filter are plotted in Fig. 7(a). The moving-average filter was applied to smooth the large oscillatory signals caused by sudden changes in the inductance of the load, as well as the noise generated by the switch. Figure 7(b) shows the numerically integrated Rogowski signal without any filtering. Timing of the waveforms was offset such that all currents overlap at their rise—defined to be at 5 kA. As such, the "jitter" within this context (and all further discussions of the jitter) refers to the temporal jitter in the x-ray burst inherent to the underlying physical processes and the dynamics within the ball gap switch. It excludes any jitter due to the trigger. This convention was chosen because the ball gap switch was shown in another set of experiments to have an inherent jitter of the order of a few nanoseconds, whereas the external TG75 trigger was a source of large jitter that could be easily removed—for example, by using a thyratron. An additional four shots were performed with a beryllium-filtered diamond PCD located 650 mm from the X-pinch to understand the timing reliability (jitter) of the x-ray burst. The corresponding Rogowski waveforms and PCD signals are plotted in Fig. 7(c), and a close-up of the region of interest is shown in Fig. 7(d). The formation of an inductive dip—a second oscillatory structure in the $dI/dt$ plot—is a signature of a sudden change in the geometry of the load corresponding to the pinching process and an emitted x-ray burst.

An inductive dip was observed in the Rogowski waveform of every shot of the 17-shot series, with a standard deviation in timing of 10 ns. The four-shot series in Fig. 7(d) shows excellent agreement between the timing of an inductive dip and a signal on the PCD, with an average delay between the maximum x-ray signal and the inductive dip of 4.6 ns—confirming the possibility of using the inductive dip as a good quantitative diagnostic of the x-ray burst timing. Simultaneously, the four-shot series shows a standard deviation in the timing of the x-ray burst equal to 14 ns, in line with the Rogowski waveform results. A control measurement done at 80 kV with two differently filtered diamond PCDs and an AXUV SiD presented in Fig. 7(e) visually confirms the simultaneity of the inductive current dip and x-ray emission.

In 15 of the 17 back-to-back shots, image plates located around the X-pinch recorded excellent properties of the x-ray radiation emitted by the X-pinch that are summarized in Secs. IV B–IV D. It is expected that the reliability can be increased further by careful alignment of the wire load, as well as by regular surface conditioning of the electrodes. Regular bead-blasting of the transmission line surfaces was found to be key in maintaining the high reliability of the device. However, it should be noted that even with a current rise rate slower than that of conventional X-pinch drivers above the Shelkovenko condition of $dI/dt$ > 1 kA/ns, the X-pinch is reliable at imaging with 10 $\mu$m scale spatial and 1 ns temporal resolution.

## B. Hard emission

Hard x-ray emission with $E$ > 10 keV, corresponding to bremsstrahlung radiation emitted by a cascade of electrons that are accelerated across the A–K gap and hit the dense wire plasma, can be used for point-projection radiography.

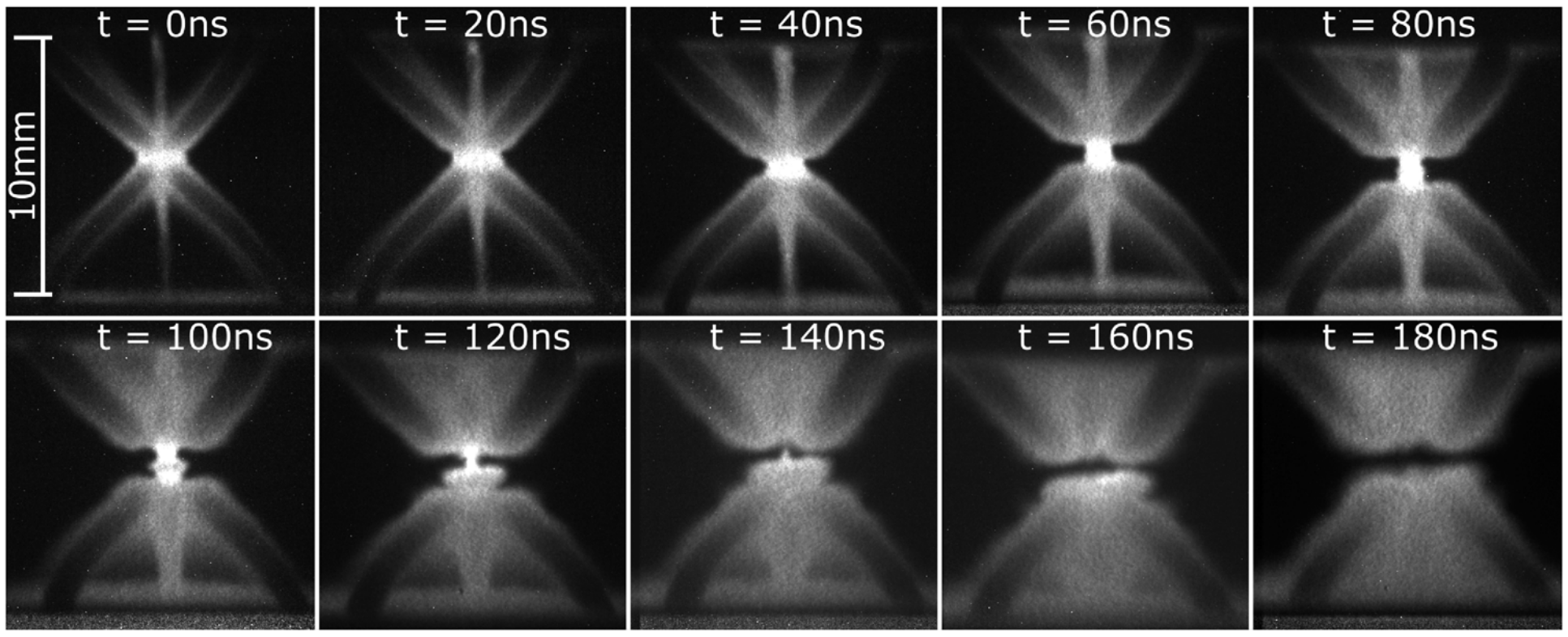

**FIG. 6.** Temporally gated optical emission emanating from the load captured every 20 ns with 5 ns exposure time. The initial frames show wire expansion followed by magnetic compression and a micropinch formed at $t$ = 120 ns. The timing of the frame showing the micropinch formation coincided with the emission recorded by the x-ray diodes to within the interframe time. The contrast is optimized for each image separately to highlight the morphology of the X-pinch.

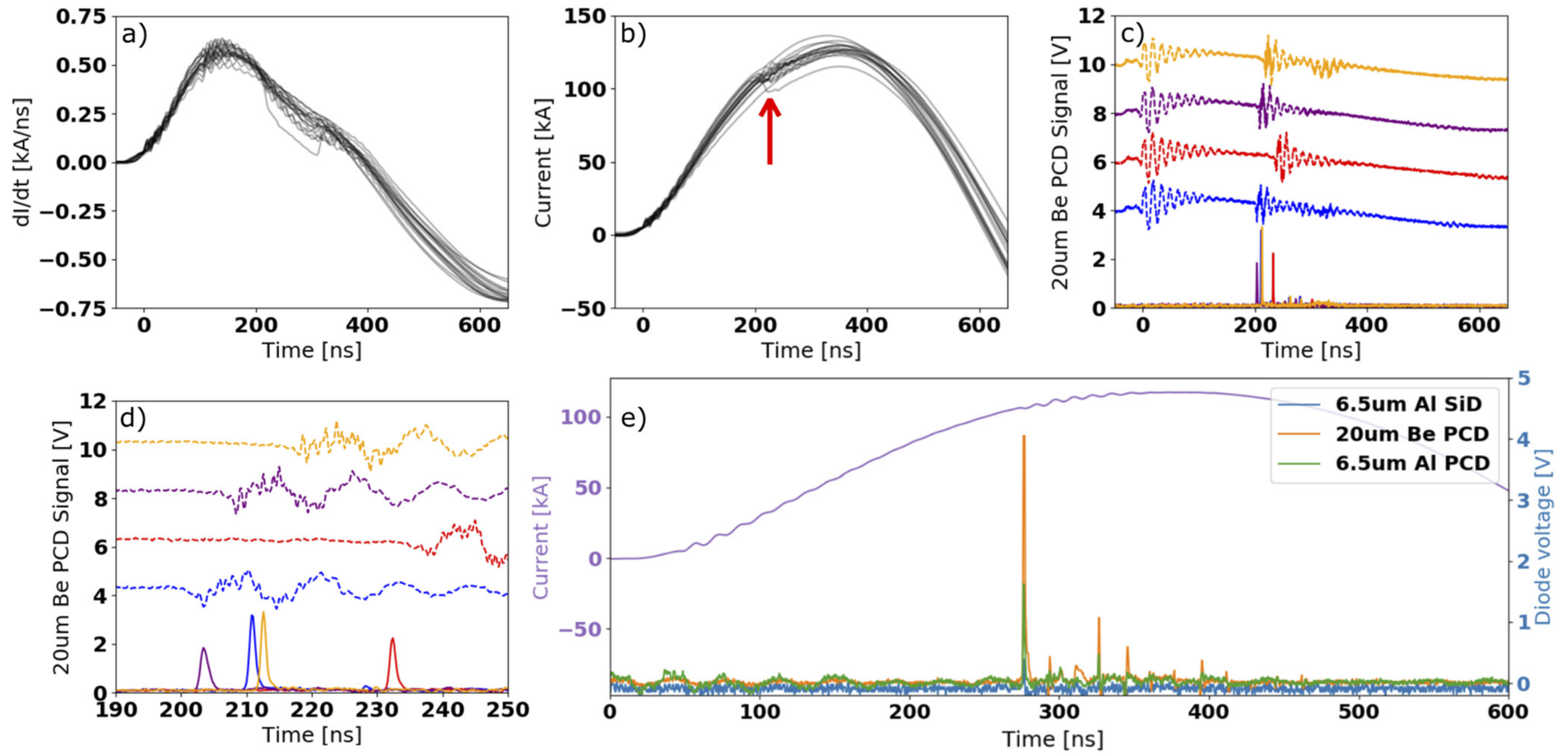

**FIG. 7.** Timing of all data in plots (a)–(d) is offset such that the current waveforms of all measurements coincide at 5 kA. (a) Current derivative time series measured by a nonintegrating Rogowski coil and smoothed by a 100-point (20 ns) moving-average filter. (b) Numerically integrated current waveforms of 17 back-to-back shots using the X-pinch driver with $4 \times 30$ $\mu m^2$ silver wire load and 85 kV charging voltage. On average, a peak current of 127 kA is achieved, with a 10%–90% rise rate of 0.45 kA/ns. The red arrow denotes the visual location of the "inductive dips." (c) The lines in the plot correspond to a signal measured by a beryllium-filtered PCD located 650 mm from the X-pinch in four separate shots (each shown in a different color). Dashed lines denote the simultaneous Rogowski coil measurement of $dI/dt$. In each of the four Rogowski waveforms, there is a second oscillatory structure—an "inductive dip" corresponding to a sudden change in the inductance during the pinching process. (d) Close-up of the PCD signals in (c). On average, the standard deviation between the timing of a maximum in the PCD signal and the onset of an inductive dip in the current waveform was 4.6 ns. (e) Example of a current waveform plotted alongside signals obtained by diamond PCDs and silicon AXUVs showing the simultaneity of the inductive dip and the x-ray emission. The FWHM of the highest x-ray burst is $\tau_{FWHM}$ = 1.1 ns as measured by all PCDs and AXUV SiDs.

Key parameters limiting the resolution of any source used for radiography are the spot size of the x-ray source, which fundamentally limits the spatial resolution, and the temporal width of the x-ray burst, which limits the temporal resolution of the measurement.

The source size of the hard x-ray hot spot was measured from point-projection radiographs of multiple known targets recorded on image plates. An example highlighting the radiographic capabilities of Dry Pinch I is shown in Fig. 8. In Fig. 8(a), a point projection of a stainless steel pinhole array covered by multiple metallic filters and located at a distance of 200 mm from the X-pinch and 250 mm in front of the image plates shows the penetrative and imaging qualities of the emitted x-rays. The emission is capable of penetrating stacked filters of 25 $\mu$m silver, 12.5 $\mu$m titanium, and 6.5 $\mu$m aluminum [transmission profiles in Fig. 5(a)] while maintaining a reasonable signal-to-background ratio of ~3. The source size of the hard x-ray emission was estimated using the sharp features in the radiograph of a USAF resolution chart in Fig. 8(b). The spatial extent of the x-ray source was measured to be 190 ± 60 $\mu$m, where the uncertainty was due to a combination of the image plate resolution and uncertainty in identification of the start and end points of transitions around the sharp features. The shot-to-shot variation in the hard source size was found to be ±100 $\mu$m. This source size was also consistent with measurements obtained from the pinhole array radiographs. A source size of 190 ± 100 $\mu$m makes the hard radiation from the X-pinch useful in radiography of large targets that do not require micrometer precision, such as impact flyers in dynamic compression experiments. An example of a common target used in such experiments is an aluminum stripline. Figure 8(c) shows a radiograph of a 1.4 mm thick aluminum stripline located at a distance of 1130 mm from the X-pinch and 1000 mm in front of an image plate. These distances were chosen as the worst-case scenario for the use of this backlighter on the M3 flyer plate accelerator—a pulsed power machine with 14 MA peak current at First Light Fusion, where Dry Pinch I will be fielded.[24]

The time duration of the hard x-ray burst was constrained using an AXUV SiD sensitive to radiation up to 50 keV. Owing to the small thickness of the sensitive element (50 $\mu$m), the effective energy cutoff of AXUVs is ~10 keV; however, their high sensitivity means that they can be used to detect much higher photon energies. There is also experimental evidence that AXUVs are highly sensitive to emission with photon energies up to 100 keV.[13] The signal recorded by the AXUV in a single measurement is plotted in Fig. 7(e). The x-ray burst measured by the AXUV coincided temporally with the soft x-ray bursts measured by the PCDs. The FWHM of the x-ray burst recorded by the aluminum-filtered AXUV was 1 ns—consistent with the PCD signals. The combination of the signals obtained by the PCDs and the AXUV imply that there is either a low amount of hard radiation or that most of the hard radiation is contained within the burst of 1 ns FWHM. The large differences between the areas under the curves of the aluminum- and beryllium-filtered PCD signals, which have different transmission coefficients between 1.5 and 10 keV, imply that most of the radiation is coupled into soft emission <10 keV.

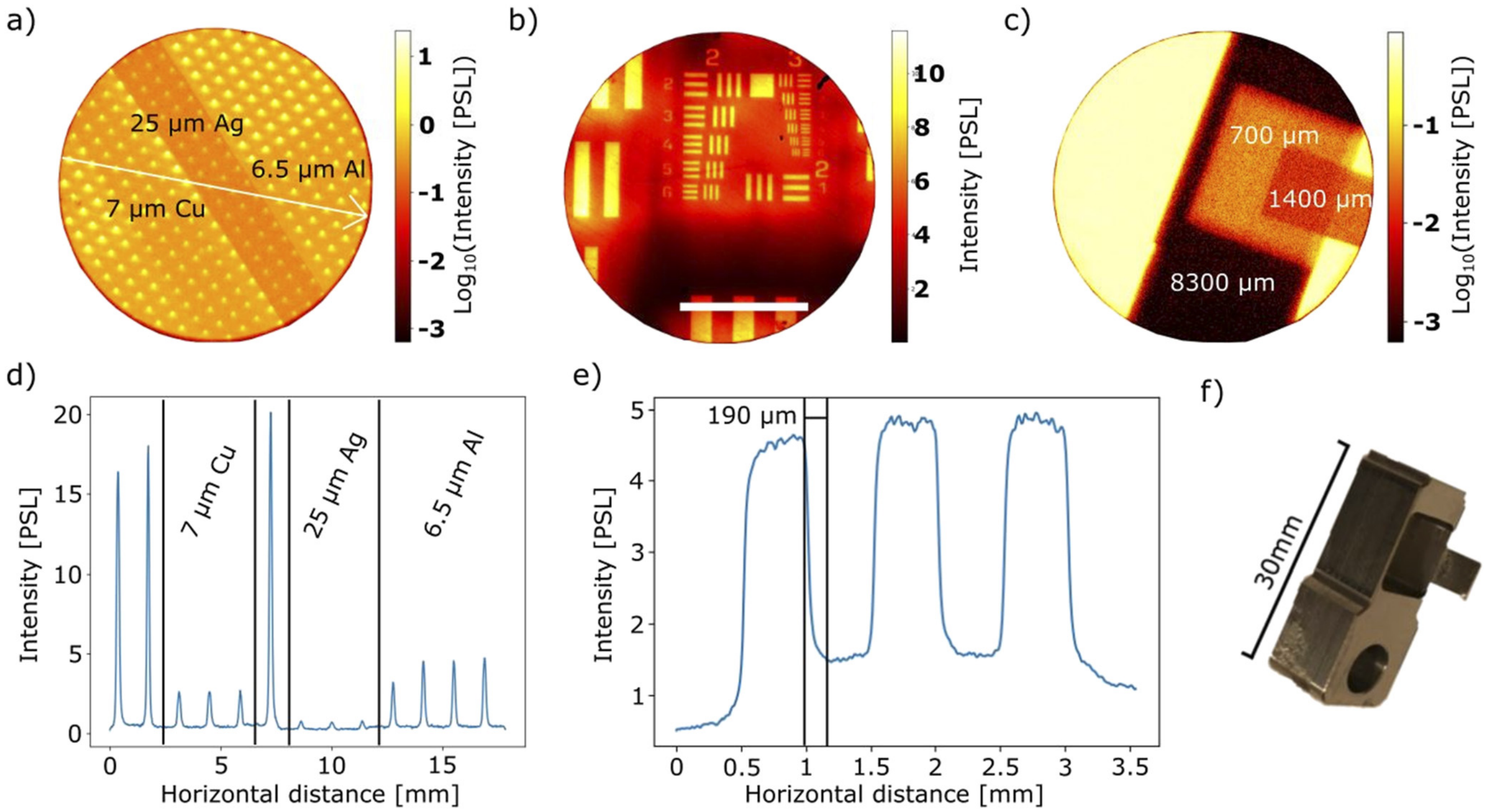

**FIG. 8.** Images of a 4 × 30 $\mu$m$^2$ aluminum X-pinch recorded on a Fuji-BAS MS image plate. All line plots are corrected for the magnification of the respective point-projection schemes. (a) Mesh made of 100 $\mu$m thick stainless steel with holes 200 $\mu$m in diameter positioned 200 mm from the X-pinch and 250 mm from the image plate. The entire mesh is covered with a 12.5 $\mu$m Ti filter with three additional filters glued to the mesh: 6.5 $\mu$m Al covering the right half of the mesh, and two strips of 25 $\mu$m Ag and 7 $\mu$m Cu covering the top and bottom. (b) Image of a USAF resolution chart positioned 210 mm from the X-pinch and 920 mm from the image plate covered by a 12.5 $\mu$m Ti filter. (c) An aluminum flyer stripline positioned at 1130 mm from the X-pinch and 1000 mm from the image plate. (d) Line plot along the white line delineated in (a), in which the arrowhead indicates the direction of the line plot. (e) Line plot along the white line delineated across the USAF resolution chart in (b). The characteristic size of the hard x-ray source is estimated from the intensity fall-off on a sharp feature. (f) Photograph of the aluminum flyer in the same orientation as used for the point-projection radiograph shown in (c).

## C. Soft emission

Source properties of the soft x-ray emission $E < 10$ keV, corresponding to fast thermal processes happening within the X-pinch hot spot compressed to extremely high densities and pressures, were investigated using a combination of absolutely calibrated PCD signals and radiographs recorded on Kodak DEF.

The signal recorded by the diamond PCDs, presented in Fig. 7(e), shows the FWHM of the most-intense x-ray burst to be 1.1 ± 0.3 ns long, irrespective of the filter material. These measurements also agree with the average FWHM of the four successful x-ray bursts shown in Fig. 7(d) at $\tilde{\tau}_{\mathrm{FWHM}} = 1.1$ ns.

Using the absolute calibration of the PCDs as reported in the literature, and their nearly flat response up to 3 keV, the PCDs can be used to estimate the x-ray yield in the region of their sensitivity. If the soft thermal radiation emitted by the X-pinch is assumed to be isotropic, it is possible to calculate the fraction of the area covered by the 1 × 3 mm$^2$ diamond PCD placed at a distance $d$ from the X-pinch. As the soft radiation is nearly fully absorbed by the 0.5 mm thick diamond layer, it is possible to use the data from the beryllium-filtered PCD in Fig. 7(e). The total soft x-ray yield within the 1.1 ns burst is estimated to be ∼26 mJ at a charging voltage of 80 kV. This result scales to an average estimate of ∼57 mJ over the four successful measurements in Fig. 7(d) at the usual higher operating voltage of 85 kV. The large difference in the x-ray output between the two charging voltages would likely narrow with higher counting statistics.

The source size of the soft hot spot was found by recording a radiograph of a USAF resolution chart placed 210 mm away from the X-pinch and recorded on Kodak DEF 920 mm behind the target. Sharp features in the radiograph imply a source size of 10 $\mu$m in both radial and axial directions, as shown in Fig. 9.

The shot-to-shot variation in the soft x-ray hot-spot source size was found to be ∼±6 $\mu$m using data from three separate shots. A single pixel on the falling/rising edge of a sharp feature in Fig. 9 corresponded to a change in the source size measurement by 1.2 $\mu$m.

## D. Spectral properties

As a benchmark for the X-pinch radiation profile, an aluminum step-wedge with thickness ranging from 100 to 1500 $\mu$m in increments of 100 $\mu$m was placed at a distance of 200 mm from the X-pinch, with an image plate placed directly behind the step-wedge. Using the known transmission function of the step-wedge and approximating the emission spectrum as bremsstrahlung emission given by Eq. (1), it is possible to calculate a theoretical transmitted signal $S_{\text{theory}}(t)$, where $t$ is the step thickness, as an energy integral of the transmission function $T(t, E)$ multiplied by the spectrum:

$$S_{\text{theory}}(t, T, A) = \int_0^{\infty} T(t, E) I(A, E, T)\, dE. \tag{2}$$

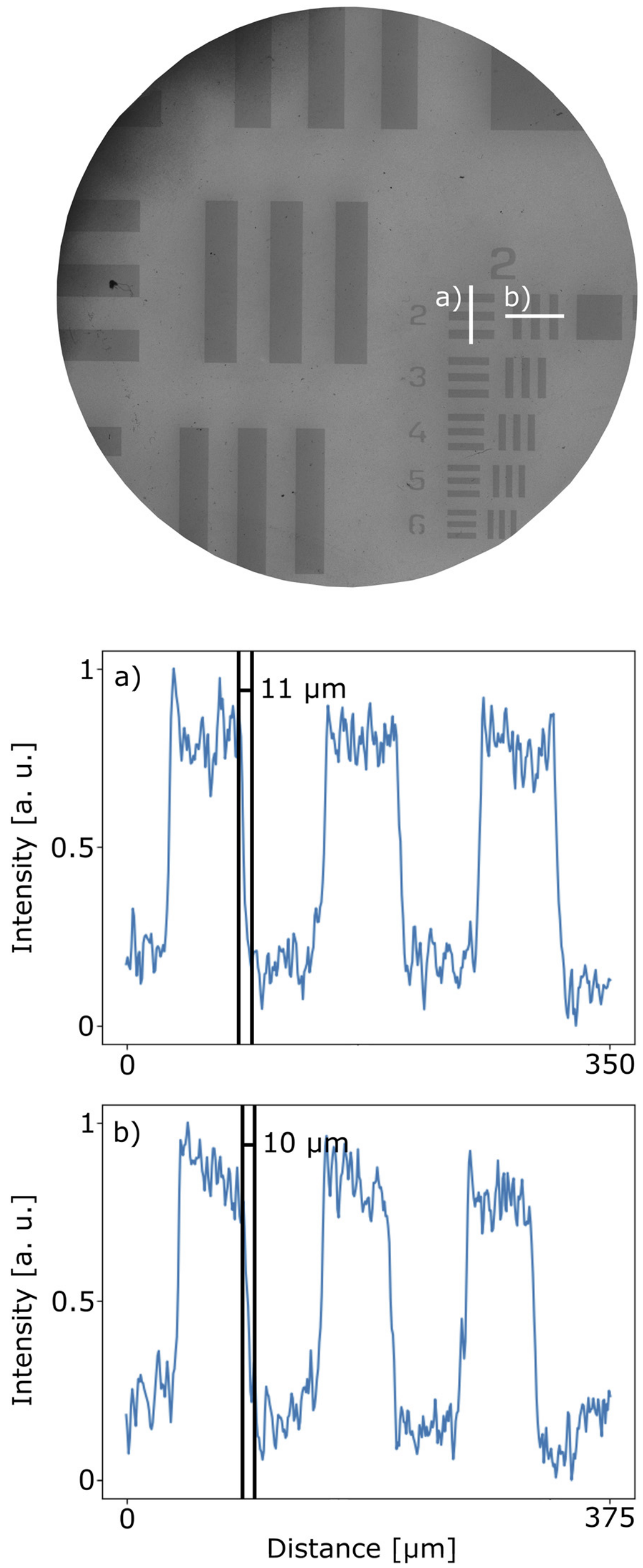

**FIG. 9.** (a) An x-ray radiograph of a USAF target covered by a 12.5 $\mu$m Ti filter positioned 210 mm from the X-pinch and 920 mm from the film plate. The image was recorded on a Kodak DEF film plate to selectively image soft x-ray radiation (see Fig. 4). (b) and (c) Vertical and horizontal lineouts of a selected element averaged over 20 neighboring bins to improve the signal-to-noise ratio. The rising and falling edges of the elements imply a soft x-ray source size of ~10 $\mu$m. The oscillatory features around the edges are caused by the graininess of the film.

Optimizing the parameters $A$ and $T$ to match the measured signal allows one to estimate a representative temperature $T$ describing the radiation "temperature" of the X-pinch. It should be noted that a temporally averaged emission does not necessarily yield a temporally averaged electron temperature. Thus, this number is not the temperature in its usual physical sense, and is used only as an empirical benchmark for the approximate profile of the X-pinch emission. A radiograph of the aluminum step-wedge and the corresponding results are shown in Fig. 10. The characteristic temperature $T$ of the emission was found via nonlinear least squares optimization to be $5870 \pm 50$ eV. All the measured data points correspond to values that are areally averaged over the entire step of uniform thickness. Error bars were set by the highest and lowest bin values included within the averages. The resulting temperature is ~2500 eV above the silver L-shell, and ~15 000 eV below the K-shell. It is therefore expected that line emission plays a significant role in the spectral profile; however, the goodness of fit of the bremsstrahlung hypothesis implies that continuum emission is the dominant source of x-ray energy.

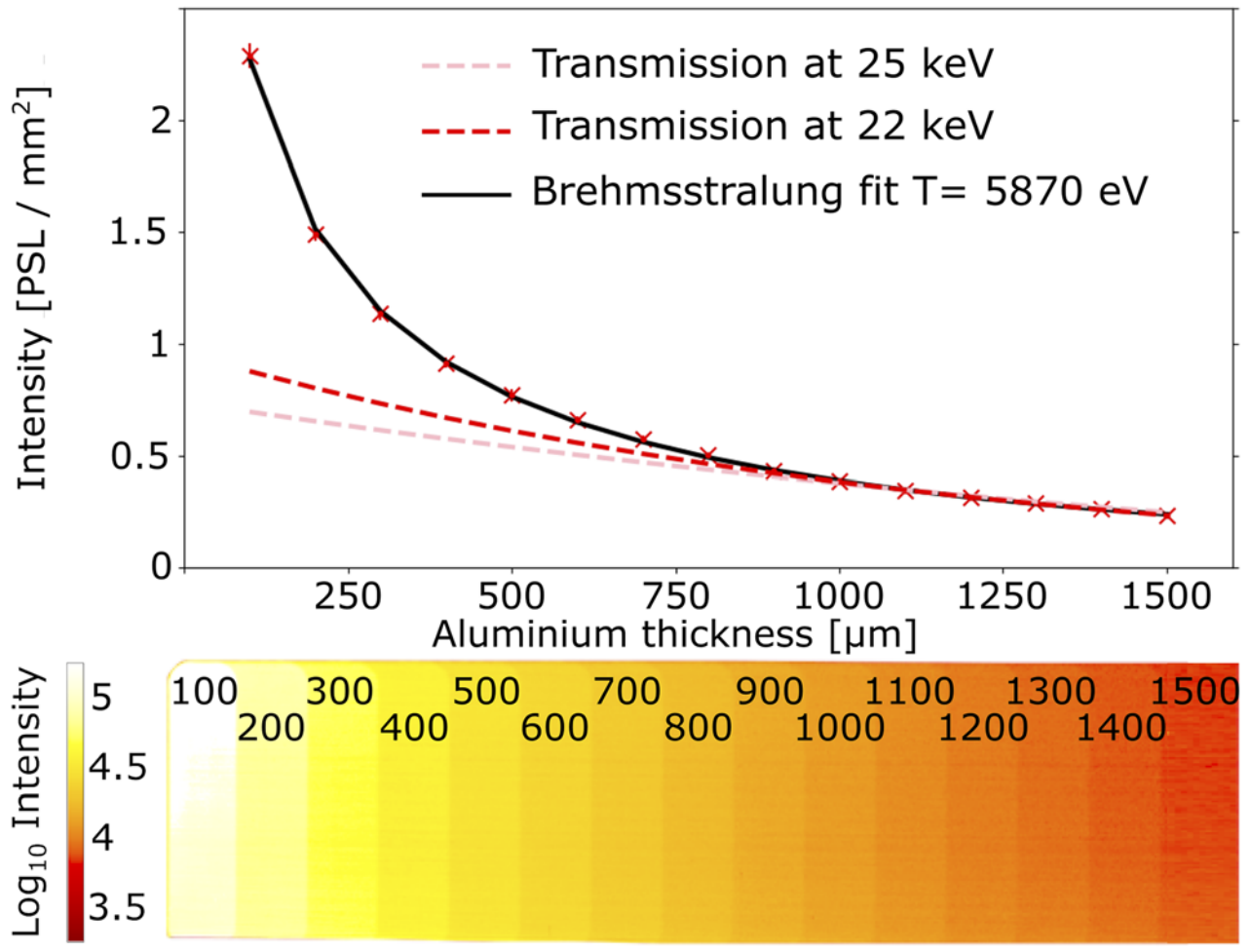

**FIG. 10.** Radiograph of an aluminum step-wedge spanning thicknesses from 100 to 1500 $\mu$m in increments of 100 $\mu$m. Signal spatially integrated over the area of respective steps is depicted by a red scatter plot. Error bars on the measured data points are calculated as the difference between the maximum and minimum pixel values in the averaging area. Optimization of a bremsstrahlung-like spectrum to match the transmitted signal yields an approximate temperature of 5870 eV. Dashed lines proportional to the transmission coefficients for energies corresponding to the K-shell emission lines of silver are included to highlight that the line emission is not the dominant process at low thicknesses, but can explain the signal at high step thicknesses.

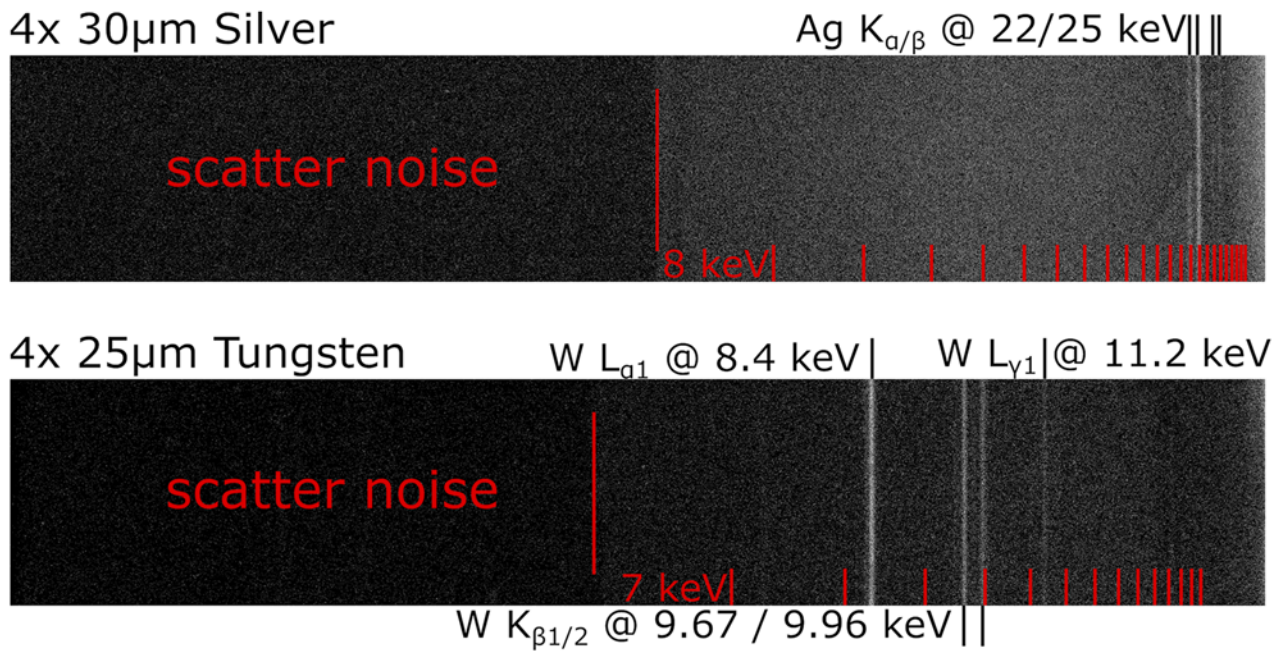

**FIG. 11.** Measurement of the X-pinch emission for silver and tungsten wire loads by a convex LiF spectrometer recorded on image plates. The LiF crystal was covered by a 12.5 $\mu$m titanium filter. The silver spectrum was obtained at a central Bragg angle of $\theta = 15^\circ$ and the tungsten spectrum at an angle of $\theta = 20^\circ$. The red line in the silver spectrum denotes the signal–noise boundary: signal is only recorded to the right of the boundary. The red markers at the bottom of the spectra denote 1 keV intervals and highlight the spatial nonlinearity of the recorded spectra.

Additional dotted lines in Fig. 10 show the theoretical transmission profiles for silver K-line-dominated emission. These curves are shown to fit the measured signal well at high thicknesses >700 $\mu$m, where x-rays with energy <10 keV do not penetrate the sample.

The spectral structure of the four-wire X-pinches was directly investigated using a convex spectrometer with a lithium fluoride crystal. Spectra for tungsten and silver four-wire X-pinches are shown in Fig. 11. These measurements support the prediction that line emission is a significant process in X-pinch radiation. Analysis of the silver X-pinch spectrum proves that the source provides mostly continuous radiation within the 7.7–36 keV region, with strong lines occurring only at higher energies, originating from the $K_{\alpha 2} = 22$ keV, $K_{\alpha 1} = 22.2$ keV, and $K_\beta = 24.9$ keV emission lines. These emission lines yielded only a weak signal, with a signal-to-background ratio of at most 2.2. In other wire materials such as tungsten, the signal-to-background ratio was as high as 10, with higher proportions of the total energy coupled into the line emission. These materials can be utilized to drive diffraction experiments as a pseudo-monochromatic source.

## V. CONCLUSIONS

Dry Pinch I is a novel X-pinch based on the pulsed power systems developed for linear transformer driver technology. It is 300 × 300 × 700 mm$^3$ in size and ~50 kg in mass, which makes it a portable and easy-to-use device that can be coupled to external experimental chambers.

In operation, it achieves peak currents of ~130 kA with an average rise rate of 0.45 kA/ns. Despite the low current rise rate, which is below the Shelkovenko condition $dI/dt > 1$ kA/ns, with proper transmission line surface conditioning, the reliability of Dry Pinch I in producing X-pinches with good imaging properties was over 88% in 17 back-to-back shots. It is expected that the reliability can be improved further by using hybrid X-pinch loads instead of four wires.

The x-ray characteristics of Dry Pinch I allow it to be a source for a wide range of diagnostics. It has been tested to work with a load consisting of four silver wires that produce hard radiation ( >10 keV) with a source size of ~200 $\mu$m that is sufficiently intense to penetrate multi-millimeter aluminum targets. The soft radiation component of its emission (1–10 keV) can be decoupled from the hard emission by stacking multiple different imaging media and has a source size of ~10 $\mu$m in both radial and axial directions. It is thus ideal for high-resolution radiography or phase contrast imaging of transparent samples. On average, the soft x-ray radiation was found to deliver ~57 mJ of energy in 1.1 ± 0.3 ns—a time resolution that allows the X-pinch to capture dynamics in high-energy-density physics experiments. The best X-pinches reported in the literature achieve x-ray bursts as short as 10 ps, and sub-micrometer spot sizes using current rise rates that exceed 1 kA/ns.[14] In the future, we will be improving our diagnostics to explore how Dry Pinch I behaves on these timescales and resolutions—at present, its output may still not be fully optimized and the hot spot formed potentially under-compressed.[25]

The spectral properties of the x-ray emission can be tuned for use in diffraction or spectroscopic measurements by using different wire materials for the load. Spectra of silver and tungsten wires were measured using a lithium fluoride spectrometer. It was shown that silver wires produce mostly continuum emission with some energy deposited into K-line emission at 22 keV. Tungsten, on the contrary, produced clear well-defined L-line emission in the vicinity of 10 keV, with little energy coupled into the continuum. These materials can be used to match the X-pinch source emission to the requirements of a particular experiment.

The shot-to-shot jitter in the timing of the x-ray burst was found to be ~10 ns, discounting any jitter caused by the trigger, making it realistic to couple the X-pinch to external experiments. In the future, Dry Pinch I is planned to drive x-ray diagnostics—such as x-ray absorption spectroscopy—for experimental campaigns on larger pulsed power systems such as MAGPIE.[26] Future generations of the Dry Pinch system are expected to include fully internal charging and safety systems, with no necessity for external components and with all controls being handled through an Ethernet cable connected to a personal computer.

## ACKNOWLEDGMENTS

Parts of the research detailed were funded by First Light Fusion Ltd., the U.K. EPSRC, and the U.S. Department of Energy under Cooperative Agreement Nos. DE-NA0003764 and DE-SC0018088.

## AUTHOR DECLARATIONS

### Conflict of Interest

The authors have no conflict of interests associated with the presented work.

## DATA AVAILABILITY

The data supporting this study are publicly available from the corresponding author upon reasonable request.